\documentclass[%
 superscriptaddress,
 twocolumn,
 amsmath,amssymb,
 aps,superscriptaddress,
 prc,longbibliography
]{revtex4-2}

\usepackage{graphicx}
\usepackage{dcolumn}
\usepackage{bm}
\usepackage{physics} 
\newcolumntype{d}[1]{D{.}{.}{#1}} 
\usepackage{xcolor}
\usepackage{soul}
\usepackage{amsthm}

\usepackage{orcidlink}
\usepackage{algorithm}
\usepackage{algorithmic}
\usepackage{booktabs}

\begin{document}


\title{Large-Scale Calculations of $\beta$-Decay Rates and Implications for $r$-Process Nucleosynthesis}

\author{A. Ravli\'c\,\orcidlink{0000-0001-9639-5382}}
\email{ravlic@frib.msu.edu}
\affiliation{Facility for Rare Isotope Beams, Michigan State University, East Lansing, Michigan 48824, USA}
\affiliation{Department of Physics, Faculty of Science, University of Zagreb, Bijeni\v cka c. 32, 10000 Zagreb, Croatia}

\author{Y. Saito\,\orcidlink{0000-0003-1320-8903}}
\email{saitoy@frib.msu.edu}
\affiliation{Facility for Rare Isotope Beams, Michigan State University, East Lansing, Michigan 48824, USA}
\affiliation{Department of Physics and Astronomy, University of Tennessee, Knoxville, Tennessee 37996, USA}
\affiliation{Department of Physics, University of Notre Dame, Notre Dame, IN 46556, USA}

\author{W. Nazarewicz\,\orcidlink{0000-0002-8084-7425}}
\email{witek@frib.msu.edu}
\affiliation{Facility for Rare Isotope Beams, Michigan State University, East Lansing, Michigan 48824, USA}
\affiliation{Department of Physics and Astronomy, Michigan State University, East Lansing, Michigan 48824, USA}

\date{\today}

\begin{abstract}
Nuclear $\beta$ decay is a key element of the astrophysical rapid neutron capture process ($r$-process). In this work, we present state-of-the-art global $\beta$-decay calculations based on the quantified relativistic nuclear energy density functional theory and the deformed proton-neutron quasiparticle random-phase approximation. Our analysis considers contributions from allowed and first-forbidden transitions. 
We used two point-coupling functionals with carefully calibrated time-odd terms and isoscalar pairing strength.
The new calculations display consistent results for both employed functionals, especially near the neutron drip line, suggesting slower $\beta$ decays past the $N=126$ neutron shell closure than in commonly used $\beta$-decay models. The new rates, along with the existing rates based on the recent non-relativistic global calculations, are found to slow down the synthesis of heavy elements in the $r$-process and significantly reduce the contribution of neutron-induced fission.
\end{abstract}

\maketitle

\section{Introduction \label{sec:intro}}
The rapid neutron capture process ($r$-process) is responsible for synthesizing the heaviest elements that naturally exist in the Universe \citep{BBFH, Cameron, Horowitz2019, Cowan2021, Arcones2022}. Although its exact site remains elusive, it is expected to occur in explosive and neutron-rich astrophysical events, such as compact binary mergers and rare types of core-collapse supernovae. The process involves thousands of species of neutron-rich nuclei, many of which have yet to be experimentally studied. They undergo various nuclear reactions and decays, including neutron capture, photodissociation, $\beta$ decay, and fission. As most of the nuclei involved are beyond the current reach of experiments, theoretical description of their properties has a profound impact on our understanding of the $r$-process. As a step towards consistent evaluation of the required nuclear physics input with controlled extrapolation, this work focuses on the influence of $\beta$ decay. The basic role of $\beta$ decay in the $r$-process is to transfer the material from the isotopic chain with proton number $Z$ to the one with $Z+1$. During the $(n,\gamma) \leftrightarrow (\gamma, n)$ equilibrium, the most abundant isotope in a given isotopic chain, so-called the ``waiting-point'' nucleus, effectively determines the $\beta$-decay rate of the isotopic chain. The half-lives of waiting-point nuclei with neutron magic numbers $N=82$ and $126$ qualitatively explain the origin of the second and third $r$-process abundance peaks at mass numbers $A\sim130$ and $196$, respectively. After the $(n, \gamma)\leftrightarrow(\gamma,n)$ equilibrium breaks down, $\beta$ decay, as well as $\beta$-delayed neutron emission, neutron capture, and fission, are the main mechanisms that affect the final abundance pattern 
\citep{Surman_rep_1997, Arcones_rep_2011a, Mumpower_rep_2012a, Mumpower_rep_2012b, Mumpower_rep_2012c, Goriely2013, Eichler2015}. Fission determines the end point of the $r$-process path, past $Z\sim90$~\citep{Panov2003, Martinezpinedo2007, Panov2010, Eichler2015, Giuliani2018, Mumpower2018, Vassh2019, Giuliani2020}, redistributing and increasing the relative abundance of heavy nuclei. This mechanism is referred to as ``fission (re)cycling''. Neutron-induced fission has been discussed as the dominant fission channel~\citep{Panov2005, Martinezpinedo2007, Petermann2012}; however, depending on the time scale and amount of the population of fissile nuclei, the contribution of neutron-induced fission may be significantly altered.

In the majority of $r$-process studies, $\beta$-decay data have been taken from the phenomenological Finite Range Droplet Model (FRDM), combined with the quasiparticle random phase approximation (QRPA)~\citep{Moller1997, Moller2003, Moller2019}.   More microscopic approaches, whether based on the nuclear shell model~\citep {Sharma2022,DeGregorio2024} or ab initio methods~\citep{Gysbers2019,Li2025}, remain limited due to the unfavorable scaling of problem complexity with system size. Leveraging the balance between scalability and rigorous microscopic description, the energy density functional (EDF) theory is a suitable many-body method for providing reliable predictions of nuclear properties for $r$-process studies, as it is computationally tractable across the entire chart of nuclides. Starting from the nuclear ground state, excitations can be obtained by considering the linear response of the nucleus within QRPA. Extending the model space beyond two-quasiparticle configurations included in QRPA, although leading to improved agreement with experiment, remains numerically challenging and was applied to specific nuclei only~\cite{Robin2016,Robin2018a,Liu2024a}. Until recently, most large-scale calculations of $\beta$-decay rates based on the nuclear EDF framework and QRPA have been restricted to spherical nuclei \citep{Engel1999a, Niu2013, Marketin_RQRPA_2016}. Only a single global calculation has considered axially-deformed nuclei~\citep{Ney2020}. 
Since global sets of theoretical $\beta$-decay rates are scarce, the literature on the sensitivity of $r$-process simulations on $\beta$ decay has been limited \citep{Niu2013, Mumpower2014, Lund2023, Kullmann2023, Hao2023}, with the primary focus on the $r$-process observables such as the abundance pattern and electromagnetic emission. The dynamic effect of $\beta$ decay on the evolution of other reactions, decays, and abundances has rarely been explicitly discussed. In particular, the impact of $\beta$-decay modeling on the fission contribution has been largely unexplored.

In this work, we present new global $\beta$-decay calculations using the recently developed axially-deformed relativistic QRPA framework \cite{Ravlic2024a}, with the residual interaction optimized in Ref. \cite{Ravlic2025a}. Using this tool, we investigate the effect of the $\beta$  decay on the $r$-process dynamics, through comparisons with other global  models. 

This paper is outlined as follows. In Sec. \ref{sec:method} we describe theoretical methods for both nuclear $\beta$ decay and $r$-process calculations. Section \ref{sec:results} presents benchmarks of $\beta$-decay rates with experimental data and other theoretical calculations, discussions on $Q_\beta$-values and impact of first-forbidden transitions, as well as quantified impact of the new rates on different $r$-process scenarios. Conclusions and an outlook are given in Sec.~\ref{sec:conclusions}. All results compiled in this work are available in a public repository~\cite{GitHub}, together with an online interactive visualization tool to display observables across the nuclide chart~\cite{rpx_website}.

\section{Theoretical Methods \label{sec:method}}

\subsection{Calculation of $\beta$-decay half-lives}

The ground-state of the nucleus was obtained with the relativistic Hartree-Bogoliubov theory assuming axially-deformed and reflection-invariant nuclear shapes~\citep{Niksic2014a}. Two point-coupling interactions were employed in this work: DD-PC1~\citep{Niksic2008a} and DD-PCX~\citep{Yuksel2019a}, both adopting  separable pairing interaction~\citep{Tian2009}. The ground-state quadrupole deformation was obtained by performing constrained calculations with the quadrupole deformation  $\beta_2 \in [-0.5, 0.5]$, and subsequently releasing the constraint to find the configuration that minimizes the total energy. Odd-$A$ and odd-odd nuclei were treated by self-consistent quasiparticle blocking within the equal-filling approximation (EFA)~\citep{Perez-Martin2008a}. Calculations were performed by discretizing quasiparticle states in a stretched axially-deformed harmonic oscillator basis. We used $N_{\rm osc} = 18$ shells for nuclei up to $Z = 103$ and for heavier nuclei $N_{\rm osc} = 20$. To obtain excited states in the daughter nucleus, we employed the linear response approach, developed in Ref.~\cite{Ravlic2024a}. The residual interaction couplings, including the isoscalar pairing interaction strength $V_0^{is}$, and Landau-Migdal coupling $g_0$, in addition to effective axial-vector coupling $g_A$, were used as optimized in the recent work~\citep{Ravlic2025a}. The $\beta$-decay rates $\lambda_\beta$ were calculated by the contour integration around a suitably chosen contour $\mathcal{C}$ encircling the $Q_\beta$ window:
\begin{equation}\label{eq:contour}
    \lambda_\beta = \frac{\text{ln} 2}{K} \frac{1}{2 \pi i} \oint_{\mathcal{C}} d\omega \int\limits_1^{W_0[\omega]} dW f(W,Z) C(W, \omega),
\end{equation}
where  the phase-space factor is
\begin{equation}
    f(W,Z) = pW(W_0[\omega] - W)^2 F(Z,W),
\end{equation}
with $W$ being the electron energy (in $m_e c^2$ unit), $p = \sqrt{W^2 - 1}$ -- its momentum (in $m_e c$), and $F(Z,W)$ being the Fermi function, taking into account the distortion of the electron wave function. The end-point energy is given by
\begin{equation}\label{eq:energy_window}
    W_0[\omega] (m_e c^2) =  \Delta_{np}-\omega + \lambda_n - \lambda_p,
\end{equation}
where $\lambda_{n(p)}$ is the neutron(proton) chemical potential, and $\Delta_{np} = 1.293$ MeV is the neutron-proton mass difference. The nuclear matrix elements are contained within the shape factor
\begin{equation}
    C(W, \omega) = k(\omega) + ka(\omega) W + \frac{kb(\omega)}{W} + kc(\omega)W^2,
\end{equation}
where $k, ka, kb,$ and $kc$ are tabulated in Refs.~\cite{Behrens1971a,Bambynek1977a}. The super-allowed decay constant was taken as $K = 6137$ s. The circular contour was discretized with a 30-point Gauss-Laguerre quadrature. The integration over electron energy $W$ was performed by Lagrange interpolation of the integrand on a 20-point Chebyshev grid, as discussed in Ref.~\cite{Ravlic2025a}. Subsequently, the half-lives were obtained as $T_{1/2} = \text{ln}2/\lambda_\beta$.  We included the allowed Gamow-Teller (GT) $1^+$ transitions, as well as the first-forbidden (FF) $0^-, 1^-$, and $2^-$ $\beta$ decays. Calculations were performed in the range of nuclei  between  $Z = 20$ and  $Z = 110$, containing 4598 individual nuclides. For each isotopic chain, calculations started close to the valley of stability, and ended at the two-neutron drip line  theoretically estimated  in Ref.~\cite{Ravlic2023a}.

\begin{table*}[!hbt]
\centering
\caption{Summary of global $\beta$-decay models considered in this work for $r$-process simulations, including the  method used to compute the rates, restrictions on nuclear shapes, and contribution of different multipoles. GT corresponds to $1^+$ and FF to $0^-, 1^-$ and $2^-$ multipoles.}
\label{tab:beta_models}
\renewcommand{\arraystretch}{1.15}
\begin{tabular}{@{}l l l l@{}}
\toprule
\textbf{Model} &
\textbf{Method} &
\textbf{Geometry} &
\textbf{Multipoles} \\
\midrule
FRDM \cite{Moller2019} &
micro-macro + QRPA({schematic}) &
deformed  &
GT+FF({gross}) \\
\addlinespace
D3C$^*$ \cite{Marketin_RQRPA_2016} &
EDF(covariant) + QRPA({matrix}) &
spherical &
GT+FF \\
\addlinespace
SkO$'$ \cite{Ney2020} &
EDF(Skyrme) + QRPA({linear response})  &
deformed  &
GT+FF \\
\addlinespace
DD-PC1 and DD-PCX (this work) &
EDF(covariant) + QRPA({linear response})  &
deformed &
GT+FF \\
\bottomrule
\end{tabular}
\end{table*}

\subsection{Nuclear reaction network}
The $r$-process abundance evolution was calculated with the nuclear reaction network code \textsc{PRISM} \citep{SprouseTheis2020}. In addition to our new DD-PC1 and DD-PCX sets of $\beta$-decay rates, a summary of $\beta$-decay models employed in this work is shown in Table~\ref{tab:beta_models}. 
In particular, it includes the  FRDM + QRPA $\beta$-decay tables from Refs.~\cite{Moller2003, Moller2019}, based on a schematic separable QRPA interaction and phenomenological estimates of FF transitions. Other rates based on the EDF framework include the predictions from Ref.~\cite{Ney2020}, calculated with the axially-deformed charge-exchange linear response QRPA with the Skyrme SkO' functional, and relativistic matrix-QRPA results with the EDF D3C* from Ref.~\cite{Marketin_RQRPA_2016}, restricted to spherical nuclei. Other nuclear properties, such as $\beta$-delayed neutron emission probabilities, one-neutron separation energies, and fission data, are primarily based on the FRDM model and experimental data. Separate calculations were also performed with the REACLIB nuclear data library \citep{Cyburt2010}. See Sec. \ref{sec:nucleardata} for more details.

For astrophysical conditions, we first employed the parameterized wind model of Ref.~\cite{Panov2009}, with an initial entropy per baryon of $s/k_B = 40$, and an expansion timescale of 20~ms. The initial electron fraction $Y_e$, a measure of the neutron-richness of the environment, ranges from 0.15 to 0.35 in 0.025 intervals. The initial temperature of these astrophysical trajectories is 10~GK, and the initial compositions of the nuclei were calculated using the SFHo equation of state \citep{Steiner2013}. Additionally, we also used a trajectory with an initial condition with $s/k_B=20$ and $Y_e = 0.05$, to simulate a neutron-rich environment in which fission cycling is expected to occur.

To investigate the effect of the new $\beta$-decay rates in more realistic astrophysical conditions, we considered thermodynamical evolution from two sets of astrophysical simulations: a magneto-rotational supernova \citep{Reichert2021, Obergaulinger2017, Reichert_github}, in which it has been suggested that the production of nuclei past the third $r$-process peak at $A\sim195$ occurs; and outflow from the accretion disk formed after a merger of binary neutron stars \citep{Metzger2014, Holmbeck2019}. Similarly to the case of the parametrized trajectories, nucleosynthesis calculations were initiated at $T=10$~GK, and the composition at the time was calculated using the SFHo equation of state.

\subsection{\label{sec:nucleardata}Nuclear Data}
We employed two sets of nuclear data libraries for reactions and decays other than $\beta$ decay: the FRDM-based nuclear data library, which is described in the following and also in, e.g., Refs.~\cite{Holmbeck2019, saito2025}; and the REACLIB nuclear data library \citep{Cyburt2010}. To simplify the calculation, for both sets of nuclear data, we adopted symmetric fission yields without neutron emission.

In the FRDM-based nuclear data library, neutron capture rates were calculated using the statistical Hauser-Feshbach (HF) code \textsc{CoH3} with nuclear masses from FRDM2012 \citep{Moller2016} or AME2020 \citep{Wang2021} whenever experimental data are available. Photodissociation rates, the reverse reaction of neutron capture, were calculated within \textsc{PRISM} using the detailed balance based on the neutron capture rates and the one-neutron separation energies. Theoretical $\alpha$-decay rates were estimated using the Viola–Seaborg relation \citep{Viola1966} including the $Q_{\alpha}$ values based on FRDM2012 and AME2020. Whenever possible, these were replaced with experimental $\alpha$-decay rates from NUBASE2020 \citep{Nubase2020}. Spontaneous fission rates were taken from NUBASE2020. Neutron-induced fission rates were calculated with \textsc{CoH3} using the masses from FRDM2012 or AME2020 and the fission barriers from FRLDM \citep{Moller2015}. $\beta$-delayed fission rates as well as neutron-emission probabilities were calculated using the FRDM+QRPA+HF framework \citep{Mumpower2016, Mumpower2018}.

For the calculations with the REACLIB nuclear data library, most reaction and decay rates, including the photodissociation rates, were taken directly from REACLIB. However, spontaneous fission and $\beta$-delayed fission rates were taken from NUBASE2020. For the latter, the rates were supplemented by FRDM-based rates \citep{Mumpower2018}. For neutron-induced fission, we used data from Ref.~\cite{Panov2010} calculated with FRDM masses \citep{Moller1995} to maintain consistency with other REACLIB data. For both nuclear data sets, theoretical rates were replaced by the data from NUBASE2020 whenever available.

\section{Results and Discussion \label{sec:results}}

\begin{figure}[htb]
    \centering
    \includegraphics[width=0.8\linewidth]{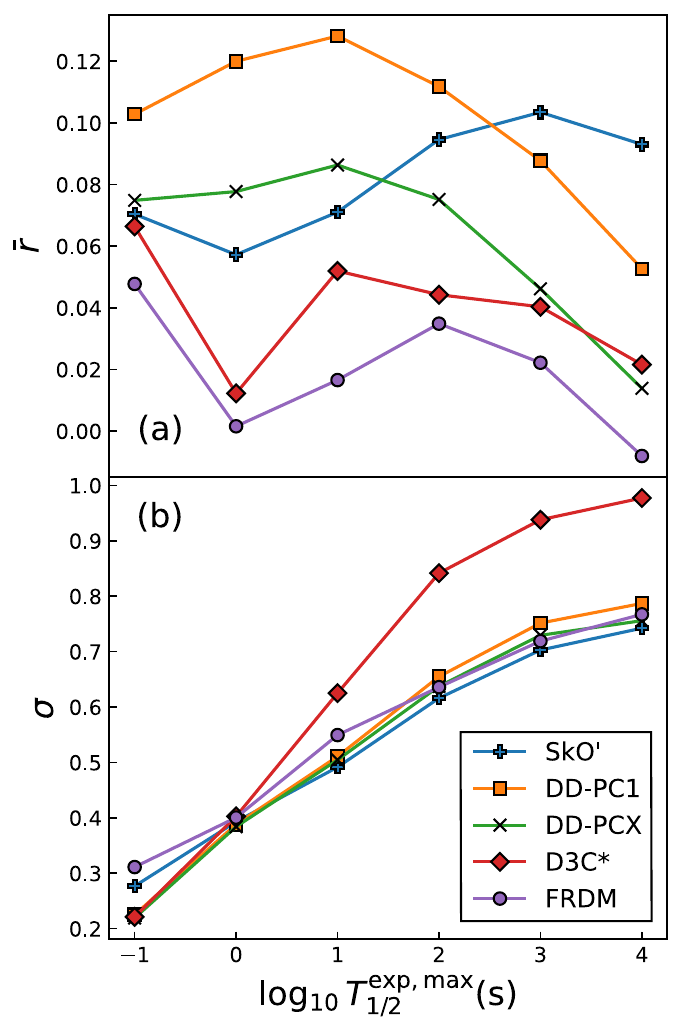}
    \caption{Global comparison between (a) mean averaged half-lives $\bar{r}$  and (b) their spread $\sigma$  obtained in this work (DD-PC1 and DD-PCX) with the results of  SkO' model~\citep{Ney2020}, D3C*~\citep{Marketin_RQRPA_2016}, and   FRDM~\citep{Moller2003}. The experimental data are taken from Ref.~\citep{Nubase2020}. Averaging in Eq. (\ref{eq:mean}) is performed over the nuclei with half-lives up to the maximum experimental half-life $T_{1/2}^{\rm exp,max}$ in the range from $10^{-1}$\,s to $10^4$\,s.}
    \label{fig:benchmark_HL}
\end{figure}

\begin{figure*}[t]
    \centering
    \includegraphics[width=\linewidth]{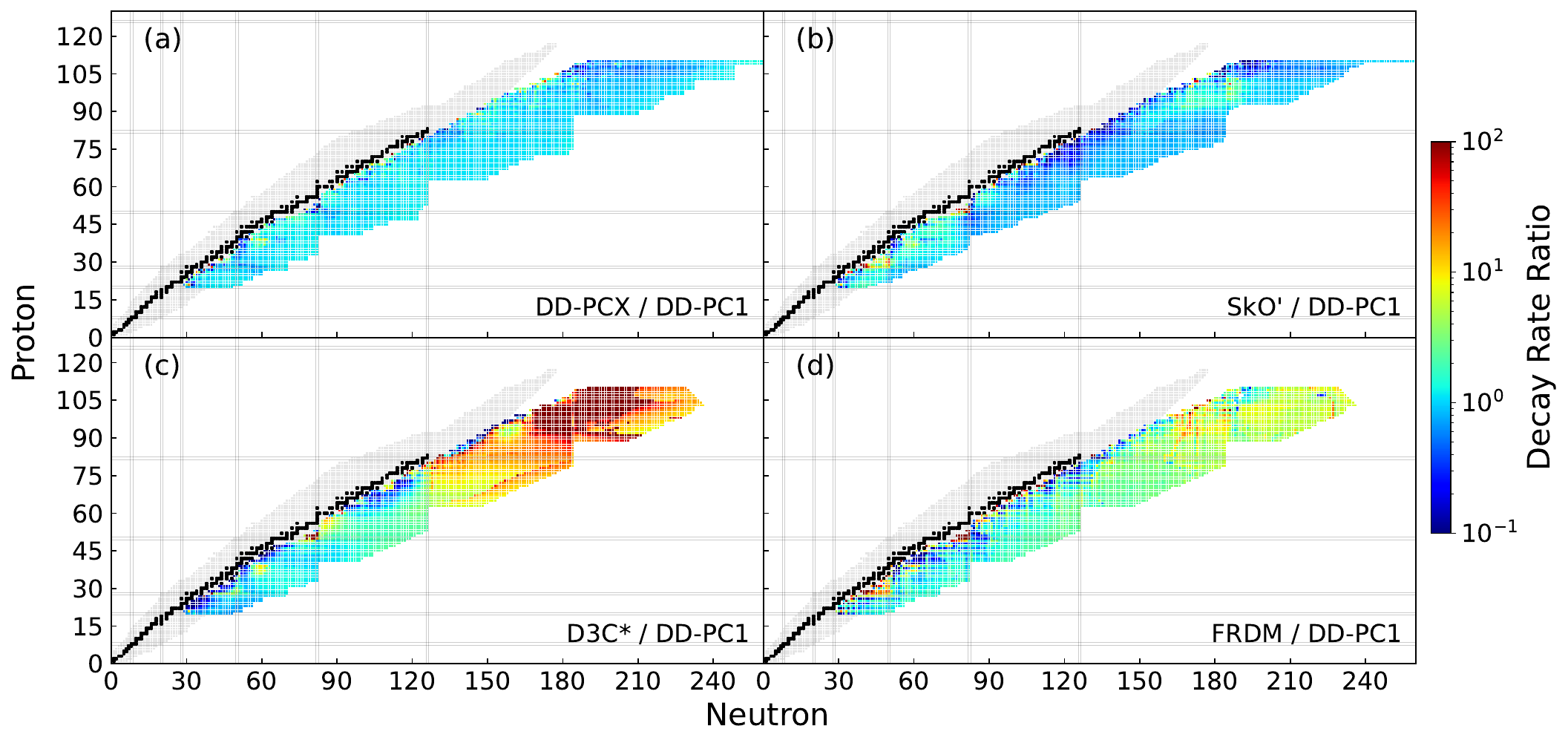}
    \caption{Comparison of theoretical $\beta$-decay rates relative to predictions of our  DD-PC1 EDF model: (a) our DD-PCX model, (b) SkO' model \citep{Ney2020}, (c) D3C* model \citep{Marketin_RQRPA_2016}, and (d) FRDM model \citep{Moller2019}.}
    \label{fig:rate_comparison}
\end{figure*}

\subsection{Benchmarking theoretical rate calculations}
As a first step in validating new rate  calculations, it is important to compare the  results  with available data from NUBASE2020~\citep{Nubase2020}. Since the half-lives span orders of magnitude, it is common to employ the following quantifiers for the mean value:
\begin{equation}\label{eq:mean}
    r_i = \log_{10}\frac{T_{1/2}^{\text{calc.}(i)}}{T_{1/2}^{\text{exp.}(i)}}, \quad \bar{r} = \frac{1}{N} \sum \limits_{i = 1}^N r_i,
\end{equation}
for a sample of $N$ nuclei with measured $T_{1/2}^{\rm exp.}$ and calculated $T_{1/2}^{\rm calc.}$  half-lives. The spread around the mean value is defined as
\begin{equation}\label{eq:sigma}
    \sigma = \sqrt{\frac{1}{N} \sum \limits_{i = 1}^N (r_i - \bar{r})^2}.
\end{equation}

Results of the inter-model comparison are shown in Fig.~\ref{fig:benchmark_HL}, by restricting the summations in Eqs. (\ref{eq:mean}) and (\ref{eq:sigma}) to nuclei with experimental half-lives below $T_{1/2}^{\rm exp,max}$, i.e., $T_{1/2}^{\text{exp.}(i)} < T_{1/2}^{\rm exp,max}$. Our DD-PC1 and DD-PCX results are compared to SkO'~\citep{Ney2020}, D3C*~\cite{Marketin_RQRPA_2016}, and FRDM~\citep{Moller2003,Moller2019} predictions. In this comparison, we only consider nuclei with $T_{1/2}^{\rm calc.} < 10^6$\,s. As shown in Fig.~\ref{fig:benchmark_HL}(a), all models tend to overestimate the experimental half-lives. Both DD-PC1 and DD-PCX show a similar pattern, with the maximum discrepancy around $T_{1/2}^{\rm exp,max}=10$\,s. In particular, DD-PCX shows excellent accuracy at $T_{1/2}^{\rm exp,max}=10^4$~s, with the mean relative error under 5\%. For nuclei with shorter experimental half-lives, the maximum discrepancy is 20\% for DD-PCX and 32\% for DD-PC1. Overall, the mean error of our models is similar to other global theoretical models. In Fig.~\ref{fig:benchmark_HL}(b), we show the spread of the predictions around the mean, Eq. \eqref{eq:sigma}. Here, all models display a universal trend where 
$\sigma$  gradually increases with  $T_{1/2}^{\rm exp, max}$.

\begin{figure*}
    \centering
    \includegraphics[width=\linewidth]{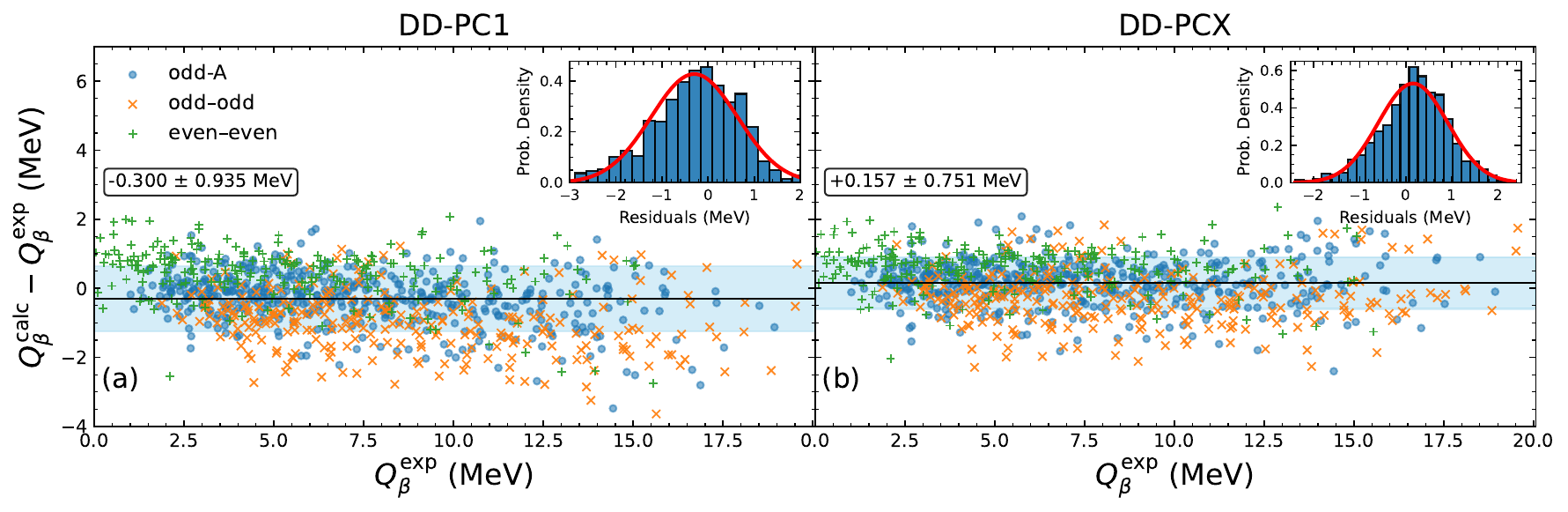}
    \caption{The residuals, $Q_\beta^{\rm calc} - Q_\beta^{\rm exp}$, between calculated and experimental $Q_\beta$-values for DD-PC1 (a) and DD-PCX (b) EDFs, as a function of experimental $Q_\beta^{\rm exp}$-values taken from AME2020~\cite{Wang2021}. The dataset is split into even-even (+), odd-$A$ 
    (circles), and odd-odd (x) nuclei. The solid  line indicates the mean value, and the band shows the $1\sigma$ interval. Distribution of the residuals, along with the normal distribution fitted to the data, is shown in insets.} 
    \label{fig:Q_values}
\end{figure*}

In Fig.~\ref{fig:rate_comparison} we compare different theoretical $\beta$-decay rates, across the nuclear landscape relative to DD-PC1. Overall, both EDFs from this work show excellent agreement with SkO'. Discrepancies are seen closer to the valley of stability, where half-lives are quite long, meaning that the rates are determined by only a few peaks of the strength function within the $Q_\beta$ window. Otherwise, regions of somewhat larger discrepancies exist around magic nuclei with  $N = 50, Z = 28$, $N = 82, Z = 50$, and $N = 184$. While our rates tend to be smaller than SkO' for these nuclei, the differences are within an order of magnitude. The excellent agreement with SkO' towards the neutron drip line is particularly remarkable, considering significant model extrapolations involved. Additionally, both EDFs used in this work show very consistent predictions, indicating a relatively small systematic uncertainty due to the choice of a functional. On the other hand, the D3C* spherical model exhibits significant discrepancies for $Z > 60$ and $N > 126$. We note that it is difficult to assess whether the differences solely originate from the assumption of sphericity in D3C* or  due to the specifics of effective interaction. Somewhat smaller differences are obtained for FRDM, around one order of magnitude, being especially pronounced around and above the $Z = 82$ shell closure.
The half-lives predicted in this work, $Q_\beta$ values, and other ground-state properties are available in a public repository~\citep{GitHub}.

\subsection{$Q_\beta$-value calculations}
The $Q_\beta$-values are calculated by determining the relativistic Hartree-Bogoliubov ground states in the parent and daughter nuclei:
\begin{equation}
    Q_\beta^{\rm calc} = B(Z,N) - B(Z+1, N-1) + \Delta_{nH},
\end{equation}
where $B(Z,N) < 0$ is the calculated binding energy and $\Delta_{nH} = 0.782$ MeV is the neutron-hydrogen atom mass difference. Calculations of $Q_\beta$ values are challenging from a theoretical perspective because they require explicit considerations of odd-$A$ and odd-odd nuclei, which are  often approximated by considering one- and two-quasiparticle excitations on top of even-even vacuum \cite{Engel1999a}. In this work, we explicitly calculate binding energies of odd-$A$ and odd-odd nuclei within the EFA blocking procedure.

Figure \ref{fig:Q_values}  shows the residuals between the calculated and experimental $Q_\beta$ values, $Q_\beta^{\rm calc} - Q_\beta^{\rm exp}$, as a function of $Q_\beta^{\rm exp}$, for DD-PC1 (a) and DD-PCX (b) EDFs for 959 nuclei for which the experimental data are available~\cite{Wang2021}. The dataset is split into even-even, odd-$A$ and odd-odd parent nuclei. The mean value of the  dataset and the $1\sigma$ deviation are marked. Regarding the global averages, the DD-PC1 calculations tend to slightly underestimate the experimental results with $\mu = -0.300$ MeV, whereas the DD-PCX calculations slightly overestimate the experiment with $\mu = 0.157$ MeV. The $1\sigma$ discrepancy is also lower for DD-PCX by around 0.2 MeV, and it visibly produces a narrower distribution around the mean in Fig. \ref{fig:Q_values}(b). It is important to note that different classes of nuclei have different contributions to the total error. Since calculations of even-even or odd-odd nuclei require both proton and neutron blocking, either in the parent or in the  daughter nucleus, it is to be expected that they are prone to  larger errors. In particular, the mean value on the dataset of 233 even-even nuclei is $\mu = 0.431(0.613)$ MeV for DD-PC1(DD-PCX) EDF, while for 260 odd-odd nuclei it is $\mu = -0.977(-0.280)$ MeV. On the other hand, odd-$A$ nuclei show the smallest mean deviations, being $\mu = -0.288(+0.173)$ MeV for DD-PC1(DD-PCX) EDF. The $1\sigma$ deviation for all the datasets does not exceed 1\,MeV. Certainly, further improvements to EDFs, and especially concerning the isovector properties, are required to improve the accuracy of $Q_\beta$-values. 

The insets in Fig.~\ref{fig:Q_values} show the distribution of the residuals. For both EDFs, the distributions are nearly-Gaussian, which is  indicative of the absence of obvious systematic effects.

\begin{figure}
    \centering
    \includegraphics[width=\linewidth]{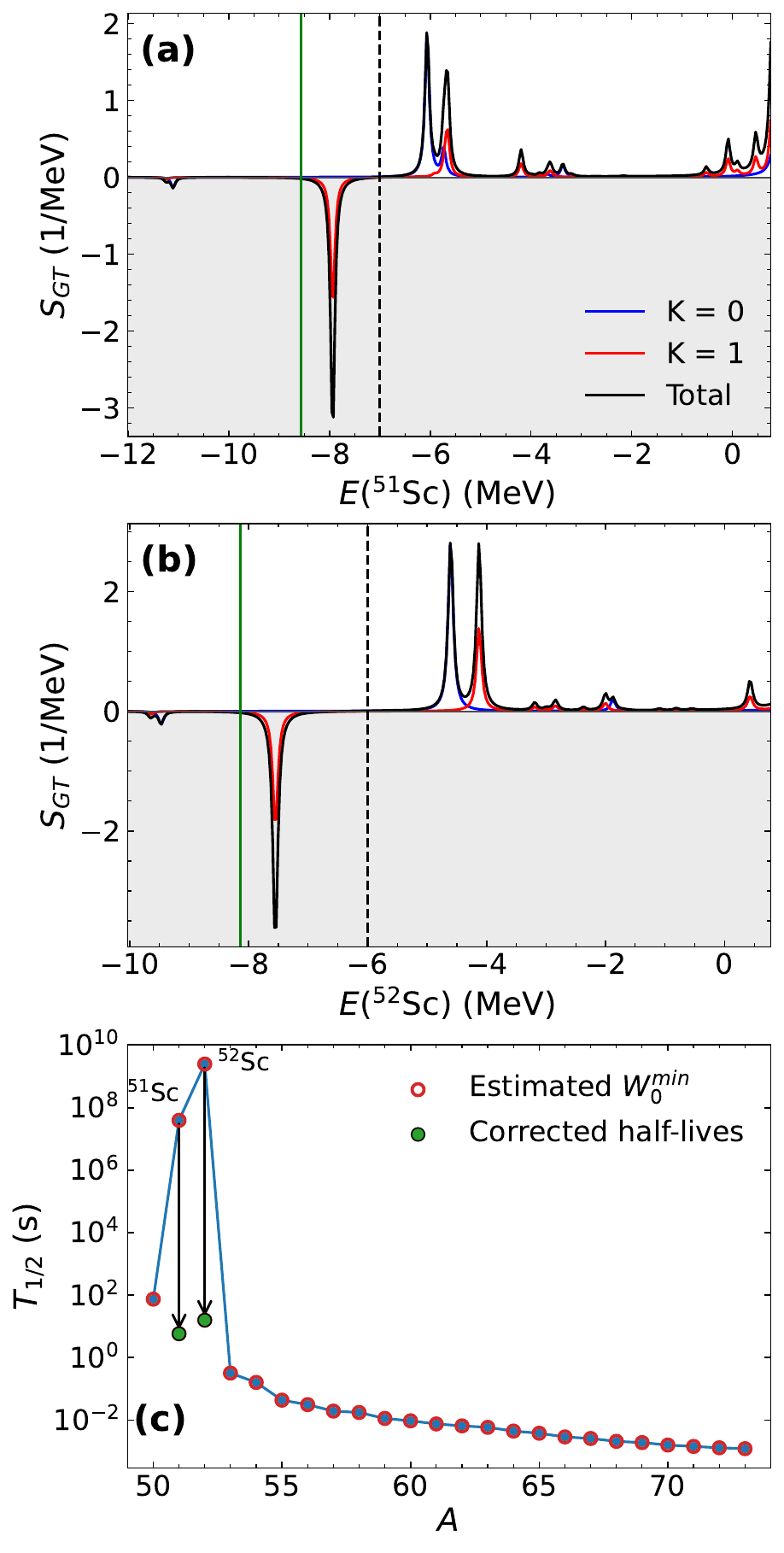}
    \caption{Correction procedure for half-lives to eliminate the spurious contribution of the negative strength for odd-$A$ and odd-odd nuclei. Panels (a) and (b) show the GT strength, decomposed into $K = 0$ and $K = 1$ components, together with their sum (Total) in ${}^{51}$Sc (a) and ${}^{52}$Sc (b). The vertical dashed line marks the manual correction made to remove the  negative strength contribution, while the solid line denotes the limit (\ref{eq:estimate}). The result of the  correction is  shown in panel (c), resulting in a significant decrease  of ${}^{51}$Sc and ${}^{52}$Sc half-lives. Calculations are performed by employing the DD-PC1 interaction.}
    \label{fig:Sc_correction}
\end{figure}

\subsection{Negative strength correction }
As demonstrated in Ref.~\cite{Ney2020},
contributions from $\beta^+$ strength can mix with $\beta^-$ strength once the EFA is employed within QRPA. The problem lies in the fact that for odd-$A$ and odd-odd nuclei, the ground-state energy of the daughter nucleus is unknown, and it has to be estimated to determine the lower end of the contour.
In particular, based on the blocking calculation, the energy window, to the leading-order, can be estimated as:
\begin{equation}\label{eq:estimate}
    W_0^{\rm max}(m_ec^2) \approx \Delta_{np} + \lambda_n - \lambda_p,
\end{equation}
as obtained from Eq.~\eqref{eq:energy_window}. Although it works reasonably well for most nuclei and transitions, the above prescription can lead to the wrong choice of integration contour in odd-$A$ and odd-odd nuclei~[cf. Eq. (\ref{eq:contour})]. Note that to obtain odd-even staggering one has to add $-\Omega_1$ term to the above equation, $\Omega_1$ being the lowest QRPA eigenvalue. A prescription to estimate this quantity was offered in Ref.~\cite{Ney2020}. However, in this work, we neglect this term for the estimation of the contour, and manually correct the calculations for nuclei with unusually long half-lives. In Fig. \ref{fig:Sc_correction} we show the GT strength function in ${}^{51}$Sc  and ${}^{52}$Sc, explicitly indicating the contour limit using the prescription from Eq.~\eqref{eq:estimate}, and then the corrected value (shift indicated by an arrow). In panel (c), we show the original half-lives calculated using the estimated window from Eq.~\eqref{eq:estimate} and the corrected values. 

To systematically correct the half-lives, first, the ``suspicious'' nuclei are identified by inspecting the individual isotopic chains. We have identified approximately 30 nuclei for both DD-PC1 and DD-PCX interactions, which necessitated these corrections. We note that such calculations are numerically expensive, since they require explicit analysis of the strength function within the $Q_\beta$-window. To achieve the best possible resolution of individual peaks in the strength function, while keeping the computational time manageable, the smearing of the Lorentzian is set at $\eta = 0.1$ MeV with energy spacing of $0.03$~MeV. Calculations are performed for the GT ($1^+$) as well as the FF ($1^-$) strength function.

\subsection{Contribution of first-forbidden transitions}

\begin{figure}
    \centering
    \includegraphics[width=\linewidth]{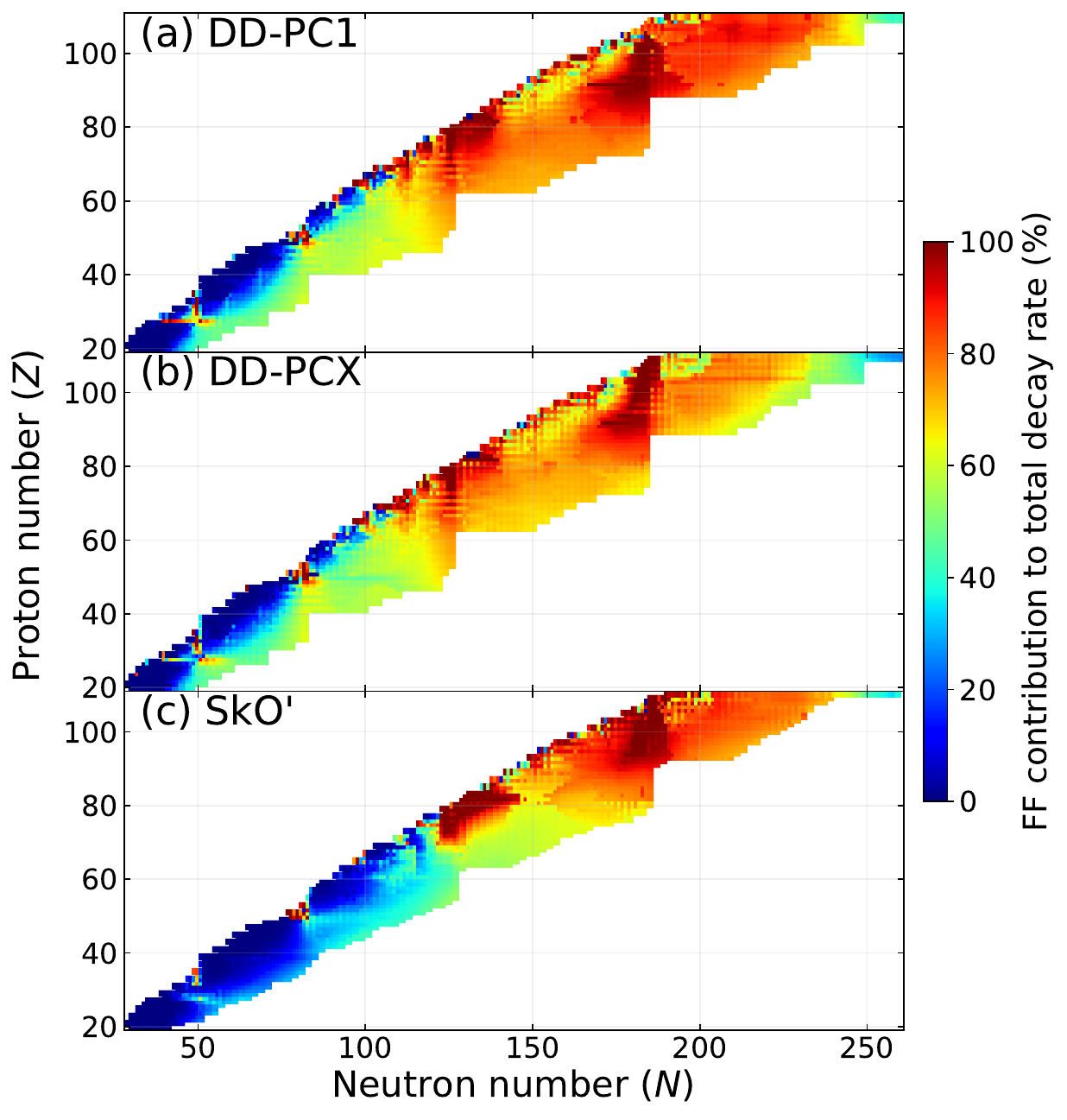}
    \caption{Contribution of first-forbidden (FF) transition (in \%) for  DD-PC1 (a) and DD-PCX (b) predictions of this work, and SkO' results (c)  from Ref.~\cite{Ney2020}.}
    \label{fig:FF}
\end{figure}

As described in Sec.~\ref{sec:method}, in our calculations, we include both allowed GT ($1^+$) and FF ($0^-, 1^-$ and $2^-$) multipoles. Detailed treatment of FF transitions, including both matrix elements and contour integration of phase-space integrals is quite involving, and can be found in Ref.~\cite{Ravlic2025a}. It is interesting to study how the relative importance of FF transitions evolves across the nuclear landscape. In Fig.~\ref{fig:FF} we compare the contribution of FF transitions to the total $\beta$-decay rate using DD-PC1 and DD-PCX models, as well as the SkO' model~\cite{Ney2020}. The universal trend in all models is that nuclei up to $Z \approx 70$ start, close to the valley of stability, with dominant contribution of GT strength to the total decay rate, which eventually decreases for more neutron-rich nuclei as the FF contribution increases. It is seen that up to $Z \approx 70$ and $N > 82$, SkO' calculations show a larger contribution of GT transitions for neutron-rich nuclei, whereas both DD-PC1 and DD-PCX estimate it to be around 40--60\%. For $Z > 70$, FF transitions tend to dominate, which is especially pronounced around the $N = 126$ and $N = 184$ shell closures, and an overall better agreement between the three models is obtained. (See also Refs.~\cite{Borzov2003,Robin2024} for the competition between allowed and first-forbidden transitions in $r$-process nuclei.)  Both models from this work, DD-PC1 and DD-PCX tend to give very consistent predictions, except for the heaviest parts of the nuclide chart beyond $Z > 90$ and $N > 200$, where DD-PC1 predicts somewhat stronger FF contribution.

\begin{figure*}[!bth]
\centering
    \includegraphics[width=0.8\linewidth]{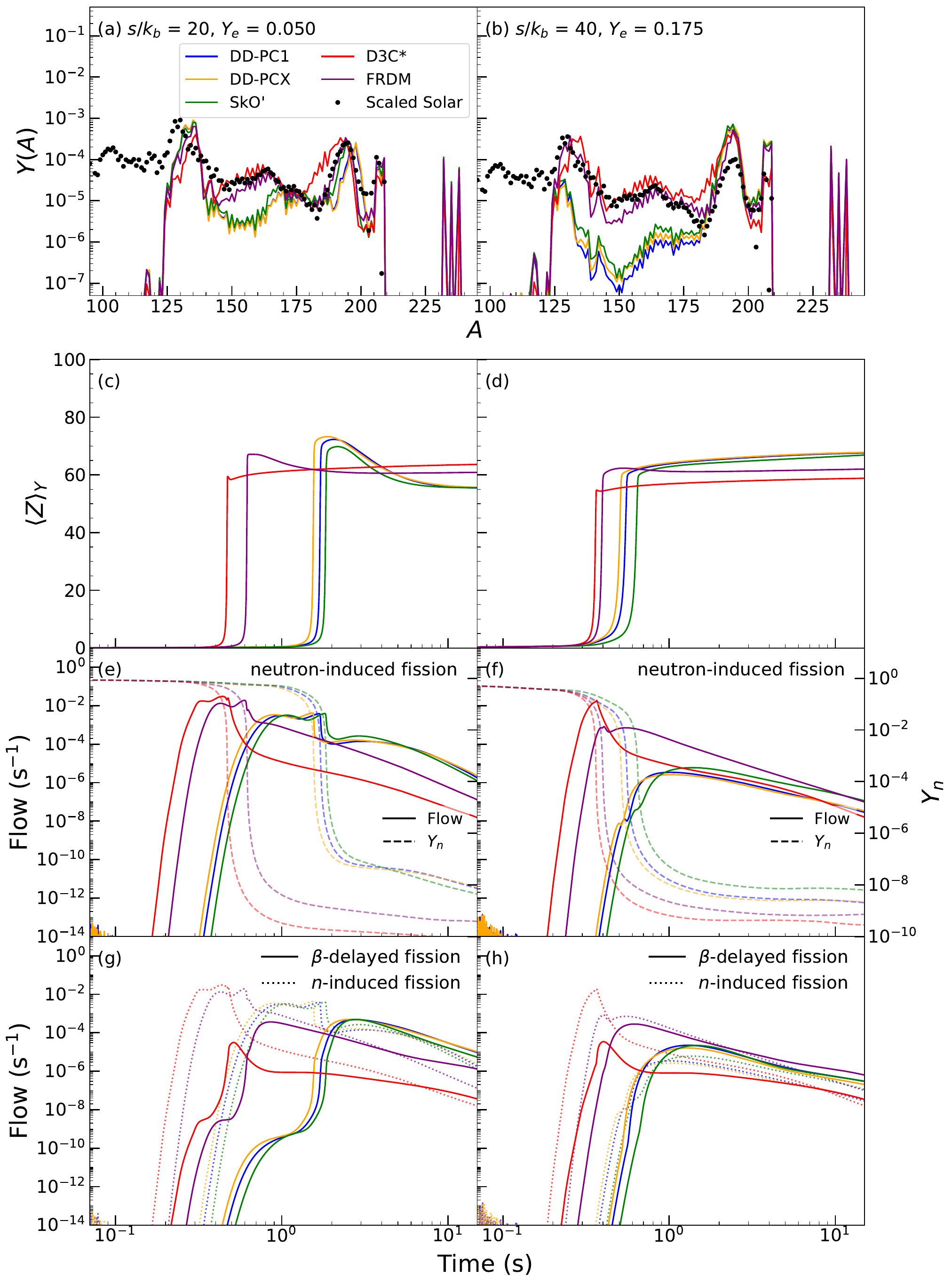}
    \caption{Final abundance patterns (a,b), the time-evolution of abundance-averaged proton number $\langle Z \rangle_Y$ (c,d), the neutron-induced fission flow and neutron abundance (e,f), and the $\beta$-delayed fission flow compared to the neutron-induced flow (g,h), for the \texttt{s20ye0050} (left column) and \texttt{s40ye0175} (right column) trajectories, respectively. The scaled solar abundance pattern from Ref.~\citep{Goriely1999_solar} is shown for reference.}
    \label{fig:parameterized}
\end{figure*}

\subsection{Impact on the \texorpdfstring{$r$}{r}-process \label{subsec:r-process}}

We have performed nuclear reaction network calculations employing the new DD-PC1 and DD-PCX values of $\beta$-decay rates and the existing predictions of SkO', FRDM, and D3C* models. Overall, we find that the $r$-process calculations using  the DD-PC1 and DD-PCX $\beta$-decay rates are generally consistent with the SkO'  calculations, and are significantly different from those with FRDM and D3C*. 

Let us define the flow $F_P$ of a process (that is, decay or reaction) $P$:
\begin{equation}
    F_P(t) \equiv \sum_{\text{all nuclei}} \lambda_P (t) \rho(t)^{N-1} \prod_{i=R_1, \ldots, R_N}Y(Z_i, A_i, t),
\end{equation}
where $\lambda_P (t)$ is the rate of the process $P$ at time $t$, $\rho(t)$ is the baryon density of the environment at time $t$, $R_1, \ldots , R_N$ are the reactants of the process $P$, and $Y(Z, A, t)$ is the abundance of the reactant $(Z,A)$. 
 
Since the results obtained with  the nuclear  properties taken from the REACLIB nuclear data library and from FRDM are very similar, in the following we will only show the results based on FRDM data in the main text. The complete set of results can be found in Ref.~\citep{GitHub}.

\subsubsection{Parameterized trajectories}
We used ten parameterized trajectories to systematically assess the effect of different $\beta$-decay predictions. Our calculations show that  $\beta$-decay rates significantly impact the abundances above the second $r$-process peak at $A\sim130$. Therefore, to highlight this effect, we focus on trajectories corresponding to neutron-rich conditions \texttt{s20ye0050} ($s/k_B=20$ and $Y_e=0.05$) and \texttt{s40ye0175} ($s/k_B=40$ and $Y_e=0.175$), where $s/k_B$ and $Y_e$ are the initial entropy per baryon and electron fraction, respectively.   \texttt{s20ye0050} approximates an extremely neutron-rich condition, such as dynamical ejecta from a neutron star merger, where fission cycling is expected to occur, whereas \texttt{s40ye0175} is a moderately neutron-rich condition. The complete set of results for ten trajectories can be found in a public repository~\cite{GitHub}.

In  Fig.~\ref{fig:parameterized}(a,b), most notable effects of $\beta$ decays  on the final abundance patterns can be seen around $140 \leq A \leq 170$ for \texttt{s20ye0050} and $130 \leq A \leq 180$ for \texttt{s40ye0175}, respectively.  Fig.~\ref{fig:parameterized}(c,d) shows that the DD-PC1, DD-PCX, and SkO' $\beta$-decay rates consistently lead to a slower increase in the average proton number, compared to FRDM or D3C*. Since the neutron density of the environment decreases as the ejecta expand over time, the delayed population of fissile nuclei also delays the fission cycling process, primarily driven by neutron-induced fission and subsequent neutron capture and $\beta$ decay. In an extremely neutron-rich condition \texttt{s20ye0050}, a significant number of neutron-induced fission events still occur, as shown in Fig.~\ref{fig:parameterized}(e). However, compared to D3C* and FRDM, subsequent neutron capture on the fission fragments would be hindered by the delayed timing of neutron-induced fission, leading to a dip in abundance patterns around $140 \leq A \leq 170$ with DD-PC1, DD-PCX, and SkO'. In the case of \texttt{s40ye0175}, Fig.~\ref{fig:parameterized}(f) shows that the delayed population of fissile nuclei leads to a significant decrease in the flow of neutron-induced fission. This results in a significant decrease in abundance below $A\sim180$ due to the absence of fission cycling. While neutron-induced fission has been considered to be the dominant channel in the $r$-process nucleosynthesis \citep{Panov2005, Martinezpinedo2007, Petermann2012}, our calculations with the DD-PC1, DD-PCX, and SkO' $\beta$-decay rates suggest that the flow of neutron-induced fission is comparable to, or only slightly larger than, that of $\beta$-delayed fission, as displayed in Fig.~\ref{fig:parameterized}(g,h). Although for the fission yields we assume a 50/50 split, as the focus of the work is the new $\beta$-decay rates, in future studies, for a more detailed assessment of the effect of fission yields, we plan to obtain realistic fission yields using the consistent EDF approach developed in Refs.~\citep{Flynn2022,Lay2024}.

\subsubsection{Trajectories from Astrophysical Simulations}
We considered two astrophysical simulations to provide more realistic $r$-process thermodynamic conditions: a mangeto-rotational supernova \citep{Obergaulinger2017, Reichert2021}, which we refer to as ``MRSN'', and accretion disk outflow from a binary neutron star merger \citep{Metzger2014, Holmbeck2019}, which we refer to as ``disk outflow''. Each case consists of a set of ten thermodynamic trajectories, and the results in select trajectories will be highlighted. The complete results of the nuclear reaction network calculations are given in Ref.~\cite{GitHub}. 

MRSN has been studied as a site that can produce elements up to the third $r$-process peak in jet-like ejecta along the axis of the rotation under strong magnetic fields, by escaping the exposure to neutrinos, which would make the ejecta less neutron-rich \citep{Mosta_MRSN_2014,Mosta_MRSN_2015,Mosta_MRSN_2018,Halevi_MRSN_2018,Cameron_MRSN_2003,Reichert_MRSN_2022,Winteler_MRSN_2012,Nishimura_MRSN_2006,Nishimura_MRSN_2015,Nishimura_MRSN_2017,Symbalisty_MRSN_1985, Reichert2021}.

\begin{figure}[htb]
    \centering
    \includegraphics[width=\linewidth]{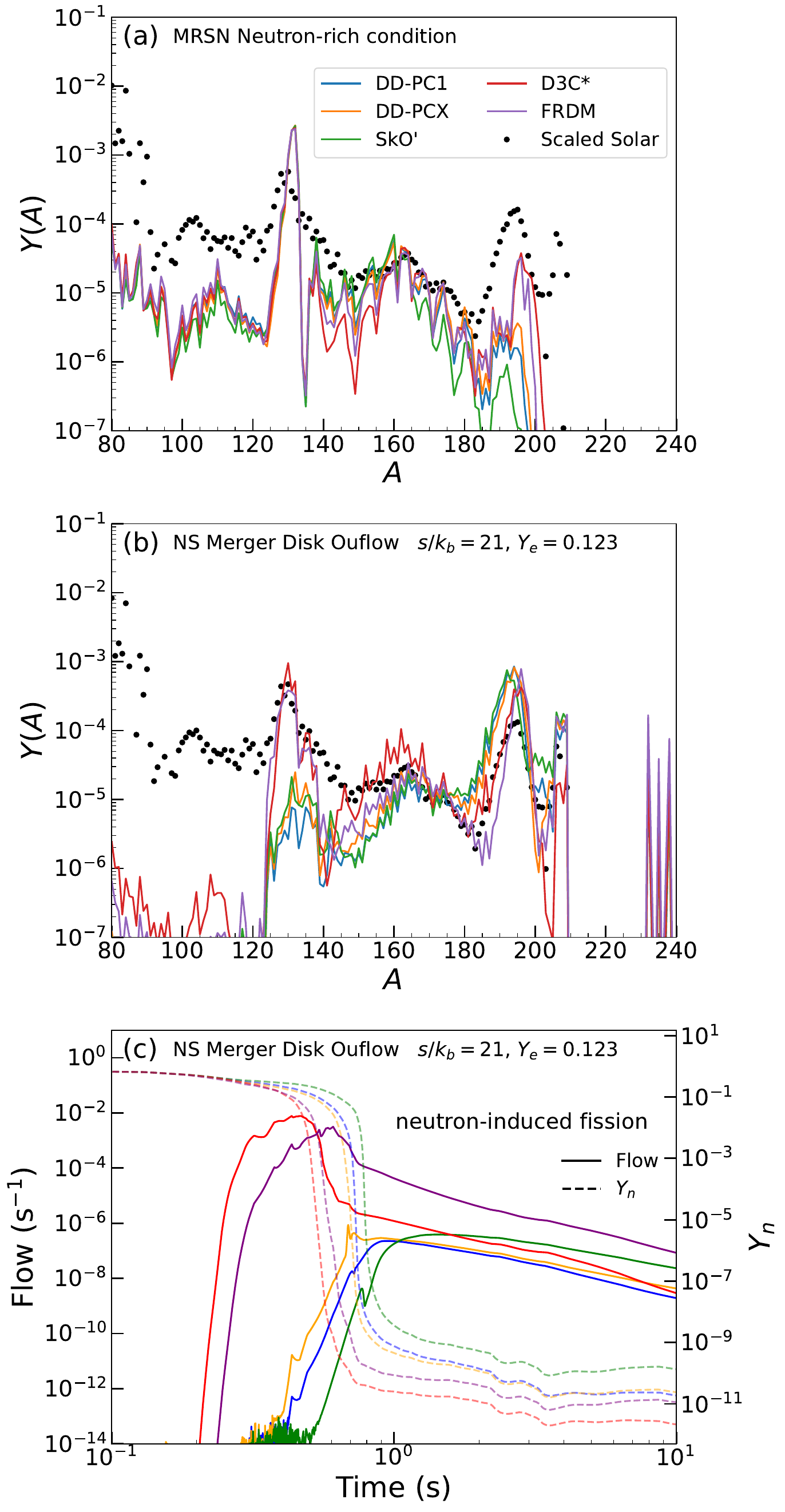}
    \caption{Final abundance patterns with different sets of $\beta$-decay rates, for (a) the most neutron-rich condition among the MRSN trajectories, and (b) the most neutron-rich condition among the disk outflow trajectories. The panel (c) shows the temporal evolution of the flow of neutron-induced fission and neutron abundance $Y_n$ for the disk outflow condition (b).}
    \label{fig:MRSN_NSMD_summary}
\end{figure}

Figure~\ref{fig:MRSN_NSMD_summary}(a) shows the final abundance patterns for the most neutron-rich condition among the MRSN trajectories, and this is the only trajectory capable of synthesizing elements up to the third $r$-process peak. With the DD-PC1, DD-PCX, and SkO' $\beta$-decay rates, the third $r$-process peak is reduced, compared to FRDM or D3C*. This indicates that $\beta$-decay rates, especially in the vicinity of and past the $N=126$ neutron shell closure, compete with the expansion timescale of the ejecta and substantially affect the nucleosynthetic outcome. The calculations with the REACLIB nuclear data library yield more pronounced third $r$-process peaks, however, DD-PC1, DD-PCX, and SkO' still consistently predict smaller peaks than FRDM and D3C*~\cite{GitHub}. These results suggest that conclusions on the production of elements in the third $r$-process peak and above in the MRSN conditions may be substantially dependent on nuclear physics inputs for the nucleosynthesis calculations, in particular $\beta$-decay rates.

For the disk outflow condition, we highlight the most neutron-rich trajectory with the initial entropy $s/k_B=21$ and electron fraction $Y_e=0.123$. The final abundances and the temporal evolution of the flow of neutron-induced fission and neutron abundance, calculated with different sets of $\beta$-decay tables combined with the FRDM-based nuclear properties, are shown in Fig.~\ref{fig:MRSN_NSMD_summary}(b,c). Similarly to the case of the parameterized trajectory \texttt{s40ye0175}, there is an orders-of-magnitude decrease in neutron-induced fission flows in DD-PC1, DD-PCX, and SkO', compared to D3C* and FRDM. With the former models, we find that the flows of $\beta$-delayed fission and neutron-induced fission are comparable. The reduced neutron-induced fission flow leads to an underproduction of the second $r$-process peak around $A\sim130$ and accumulation of abundances around $200\leq A \leq 210$.

In Ref.~\cite{Holmbeck2019}, considering similar astrophysical conditions, it was suggested that most actinide enrichments in metal-poor $r$-process enhanced stars are explained by an $r$-process source with a non-dominant amount of very neutron-rich, fission-cycling component. Given the significant effect of $\beta$ decay on fission flows indicated by our results, the understanding of actinide enrichment, in particular the distribution of the neutron-richness ($Y_e$) of the ejecta, is also likely affected by the choice of $\beta$-decay rates.
Finally, the effect of $\beta$ decay on spontaneous fission flows is also present for the parameterized and disk outflow trajectories, consistent with the findings in Ref.~\cite{Lund2023}.

\section{Conclusions \label{sec:conclusions}}
In this paper, we presented the state-of-the-art global predictions of $\beta$-decay rates, including allowed and first-forbidden transitions,    based on the relativistic nuclear EDF theory with DD-PC1 and DD-PCX functionals, including the effect of axial deformation of nuclei. The new rates agree well with the deformed non-relativistic SkO' EDF predictions~\citep{Ney2020}. These $\beta$-decay rates suggest slower $\beta$ decays, especially past the $N=126$ neutron shell closure, and are distinctly different from the phenomenological FRDM \citep{Moller2019} or spherical relativistic EDF D3C*  \citep{Marketin_RQRPA_2016} tabulations, which have been commonly used in $r$-process simulations. This significantly reduces the flow of neutron-induced fission, which affects the final abundance patterns. In less neutron-rich environments, such as magneto-rotational supernovae, the $\beta$-decay rates may compete with the expansion timescale of the ejecta, thereby affecting heavy nucleosynthesis. Since most astrophysical analyses depend on specific global $\beta$-decay models, such a choice may affect our understanding of the astrophysical environment of the $r$-process. The results presented in this work constitute a step in our effort towards comprehensive theoretical predictions based on the quantified nuclear EDF framework, which will enable a controlled assessment of uncertainties due to different nuclear properties.

\section{Acknowledgments}
We are grateful to Kyle Godbey for developing a website (\url{https://rpx.ascsn.net}) to display our results interactively. This work was supported  by the U.S. Department of Energy under Award No. DOE-DE-NA0004074 (NNSA, the Stewardship Science Academic Alliances program), the DOE Office of Science under Grants DE-SC0013365 and DE-SC0023175 (Office of Advanced Scientific Computing Research and Office of Nuclear Physics, Scientific Discovery through Advanced Computing), and by the U.S. National Science Foundation under grant number 21-16686 (NP3M).
Computational resources were provided in part by the Institute for Cyber-Enabled Research at Michigan State University. This work used TAMU ACES at TAMU through allocation PHY250120 from the Advanced Cyberinfrastructure Coordination Ecosystem: Services \& Support (ACCESS) program, which is supported by U.S. National Science Foundation grants No. 2138259, 2138286, 2138307, 2137603, and 2138296.

\bibliography{bibl}

\end{document}